\documentclass[aps, pra, showpacs, superscriptaddress , unsortedaddress,
twocolumn, preprintnumbers,]{revtex4}

\usepackage{amsmath}
\usepackage{graphicx}
\newcommand{\D}{\mbox{\rm d}}
\newcommand{\T}{\mathcal{T}}
\newcommand{\R}{\mathcal{R}}
\newcommand{\A}{\mathcal{A}}

\begin{document}
\preprint{PHYSICAL REVIEW A {\bf 74}, 033803 (2006)}
\author{A. A. Semenov}
\email[E-mail address: ]{sem@iop.kiev.ua} \affiliation{ Institut
f\"ur Physik, Universit\"{a}t Rostock, Universit\"{a}tsplatz 3,
D-18051 Rostock, Germany} \affiliation{Institute of Physics,
National Academy of Sciences of Ukraine, Prospect Nauky 46, UA-03028
Kiev, Ukraine}

\author{D. Yu. Vasylyev}
\affiliation{ Institut f\"ur Physik, Universit\"{a}t Rostock,
Universit\"{a}tsplatz 3, D-18051 Rostock, Germany}
\affiliation{Institute of Physics, National Academy of Sciences of
Ukraine, Prospect Nauky 46, UA-03028 Kiev, Ukraine}

\author{W. Vogel}
\affiliation{ Institut f\"ur Physik, Universit\"{a}t Rostock,
Universit\"{a}tsplatz 3, D-18051 Rostock, Germany}

\author{M. Khanbekyan}
\affiliation{Theoretisch-Physikalisches Institut,
Friedrich-Schiller-Universit\"{a}t Jena, Max-Wien-Platz 1, D-07743
Jena, Germany}

\author{D.-G. Welsch}
\affiliation{Theoretisch-Physikalisches Institut,
Friedrich-Schiller-Universit\"{a}t Jena, Max-Wien-Platz 1, D-07743
Jena, Germany}

\title{
Leaky cavities with unwanted noise}

\begin{abstract}
  A phenomenological approach is developed that allows one to completely
  describe the effects of unwanted noise, such as the noise associated with
  absorption and scattering, in high-$Q$ cavities. This noise is modeled by a
  block of beam splitters and an additional input-output port. The replacement
  schemes enable us to formulate appropriate quantum Langevin equations and
  input-output relations. It is demonstrated that unwanted noise renders
  it possible to combine a cavity input mode and the intracavity
  mode in a nonmonochromatic output mode. Possible applications to
  unbalanced and cascaded homodyning of the intracavity mode
  are discussed and the advantages of the latter method are shown.
\end{abstract}


\maketitle

\section{Introduction}
\label{sec1}

Cavity quantum electrodynamics (cavity QED) has been a powerful tool
in a lot of investigations dealing with fundamentals of quantum
physics and applications such as quantum information processing, for
a review see, e.\,g., Refs.~\cite{CQED, Kimble1}. It has offered a
number of proposals for quantum-state generation, manipulation, and
transfer between remote nodes in quantum networks. A cavity is a
resonatorlike device with one or more fractionally transparent
mirrors characterized by small transmission coefficients such that
large quality values $Q$ can be realized. Hence one may regard the
mode spectrum of the intracavity field as consisting of narrow
lines. As a rule, excited atoms inside the cavity serve as source of
radiation, and the fractionally transparent mirrors are used to
release radiation for further applications and to feed radiation in
the cavity in order to modify the intracavity field and thereby the
outgoing field either.

Manipulations with atoms in cavities and cavity fields give a number
of possibilities of quantum-state engineering, see, e.\,g.,
Refs.~\cite{DiFidio1, DiFidio2}. For example, schemes for the
generation of arbitrary field states have been proposed \cite{Law}.
Further, proposals for the generation of entangled-states of light
have been made (see, e.g., Ref.~\cite{Lange}). It is worth noting
that, using the technique of adiabatic atom transitions coherent
superposition states of the radiation field inside the cavity can be
prepared \cite{Parkins}.

The field escaping from an excited cavity has been proposed to be
used for homodyne detection of the intracavity mode and
reconstruction of its quantum state \cite{Santos}. Since the output
mode is a nonmonochromatic one, this proposal is based on an
operational definition of the Wigner function. The employment of
cavities as remote nodes in quantum networks has been proposed
\cite{Cirac2}. Laser driving of atoms allows one to create such a
specific pulse of the output mode which is completely coupled into
another cavity. This can be used for transferring quantum states
between spatially separated atoms trapped inside cavities. Cavities
are also important in optical parametric amplification frequently
used for the generation of squeezed states \cite{Wu}.

One of the most crucial points in implementing the proposed schemes
such as the ones mentioned above is the decoherence. It appears due
to the uncontrolled interaction of the radiation with some external
degrees of freedom giving rise to absorption and scattering of the
radiation one is interested in. In this context, a serious drawback
is the fact that for high-$Q$ optical cavities, at least with the
presently available technology, such unwanted losses can be of the
same order of magnitude as the wanted losses associated with the
transmittance of the coupling mirrors \cite{Hood, Pelton, Rempe}.
Thus, nonclassical features of the outgoing field can be
substantially reduced compared with the corresponding properties of
the intracavity field \cite{Khanbekyan}.

There exist several approaches to the theoretical description of
leaky cavities for the idealized case that unwanted losses can be
ignored. Within the framework of quantum noise theory (QNT), in
Ref.~\cite{Collett} each intracavity mode is linearly coupled,
through one or more fractionally transparent mirrors, with a
continuum of external modes forming dissipative systems for the
intracavity modes. Based on Markovian approximation, it can be
concluded that the intracavity modes obey quantum Langevin
equations. The external field is composed of two kinds of fields:
input and output ones, where the input field gives rise to the
Langevin noise forces. Moreover, the input and output fields are
related to each other by means of input-output relation.

The quantum field theoretical (QFT) approaches to the problem are
based on (macroscopic) QED. So Refs.~\cite{Knoell, Suttorp, Dutra}
start directly from an ordinary continuous-mode expansion of the
electromagnetic field in the presence of passive, nonabsorbing media
\cite{Absorb, Vogel}. Under certain conditions, this approach also
leads to a description of the cavity in terms of quantum Langevin
equations and input-output relation. In another version of the QFT
approach \cite{Viviescas}, solutions of Maxwell's equations are
constructed by using Feshbach's projection formalism
\cite{Feshbach}. Separating from the beginning all degrees of
freedom into two parts---internal and external ones---one can also
obtain, in some approximation, the Hamiltonian used in QNT.

As already mentioned, the standard versions of both the QNT approach
and the QFT approach do not take into account the presence of
unwanted losses and hence, the additional, unwanted noise
unavoidably associated with them. Thus, these theories cannot be
applied to realistic situations, in general.  Within an extended
version of the QNT approach, the unwanted losses the intracavity
field suffers from can be modeled by introduction into the Langevin
equations additional damping and noise terms, which corresponds to
the introduction into the system of additional input-output ports
\cite{Khanbekyan, Viviescas}. The applicability of this model is
restricted, in general, to the case of all the input ports being
unused.

More recently a QFT approach to the description of a leaky cavity
with unwanted noise has been presented \cite{Khanbekyan2}. Applying
quantization of the electromagnetic field in dispersing and
absorbing media \cite{scheel98, Vogel}, a generalized Langevin
equation and input-output relation have been derived. An extended
version of the QNT approach can be obtained by applying the model of
imperfect coupling between two systems, see, e.\,g., Ref.~
\cite{Gardiner}. In this scheme, a unidirectional coupling of the
considered systems is studied, the unwanted noise being modeled by
beam splitters inserted in the input and output channels.  For
transferring such a method to the description of a cavity with
unwanted noise, one must carefully check the completeness of the
parametrization of the considered replacement schemes.

In the present paper we generalize, by means of replacement schemes,
the QNT approach with the aim to complete both the quantum Langevin
equations and the input-output relation in a consistent way, such
that unwanted noise is fully included in the theory---in full
agreement with the QFT approach in Ref.~\cite{Khanbekyan2}. The
analysis will show that there exist different formulations of the
theory.  Favoring one over the other may depend on the physical
conditions and on the available information on the cavity. Moreover,
we will demonstrate that an incomplete description of the unwanted
noise may ignore important physical effects. As an example it is
shown that the unwanted noise may lead to the combination of a
cavity input mode with the intracavity mode in a nonmonochromatic
output mode. Such mode matching does not occur in an ideal leaky
cavity or in some incomplete models of unwanted noise effects.

The paper is organized as follows.  In Sec.~\ref{sec2} a
beam-splitter-based replacement scheme is introduced which is
suitable for modeling the unwanted noise of a one-sided cavity. Both
the quantum Langevin equation and the input-output relation
associated with the replacement scheme are presented. The relations
between the $c$-number coefficients in these equations are derived.
Section~\ref{sec3} is devoted to the problem of consistency and
completeness of a given quantum Langevin equation together with the
corresponding input-output relations. It is shown that the
requirement of preserving commutation rules necessarily leads to
constraints on the $c$-number coefficients in the theory. The effect
of noise-induced mode coupling between intracavity and input modes
is considered in Sec.~\ref{sec5}. Section~\ref{appl} is devoted to
the application of this effect to the problem of unbalanced and
cascaded homodyning of the intracavity mode. Finally, a summary and
some concluding remarks are given in Sec.~\ref{sec6}.

\section{
Unwanted noise
}
\label{sec2}

\begin{figure}[h!]
\includegraphics[clip=]{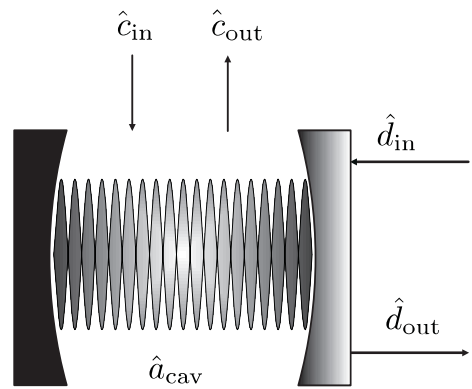}
\caption{\label{fig2} One-sided cavity with unwanted internal
losses. }
\end{figure}

As mentioned in the Introduction, the unwanted loss of cavity
photons due to scattering and absorption can be modeled by
appropriately chosen input and output ports. This is sketched in
Fig.~\ref{fig2} in the simplest case for a one-sided cavity, where
the operators $\hat{d}_\mathrm{in}(t)$ and
$\hat{d}_\mathrm{out}(t)$, respectively, correspond to the wanted
radiative input and output, whereas the operators
$\hat{c}_\mathrm{in}(t)$ and $\hat{c}_\mathrm{out}(t)$,
respectively, correspond to input and output channels associated
with unwanted noise.

The scheme
is formally equivalent
to
a
four-port
cavity having two fractionally transparent mirrors as considered
in Ref.~\cite{Collett}.
Hence any (single-mode) cavity operator $\hat{a}_\mathrm{cav}$ can be assumed
to obey
a quantum Langevin equation of the type
\begin{align}
\label{rs3}
&\dot{\hat{a}}_\mathrm{cav} (t)=  -\left[i\omega_{0}+
{\textstyle\frac{1}{2}}\left(\gamma
+\left|\A\right|^2\right)\right]\hat{a}_\mathrm{cav}(t)
\nonumber\\& \hspace{3ex}
+\sqrt{\gamma}\,\hat{d}_\mathrm{in}\left(t\right)
+\A\hat{c}_\mathrm{in}\left(t\right),
\end{align}
and the corresponding input-output relation reads as
\begin{equation}
\hat{d}_\mathrm{out}\left(t\right)
=\sqrt{\gamma}\,\hat{a}_\mathrm{cav}\left(t\right)-
\hat{d}_\mathrm{in}\left(t\right).
\label{rs4}
\end{equation}
Here, $\omega_{0}$ is the resonance frequency of the cavity,
$\gamma$ is the decay rate caused by the wanted output channel, and
$|\A|^2$ is the part of the decay rate due to unwanted internal
noise, where, for some reason which will be clarified later, $\A$ is
assumed to be a complex number. Note that such an approach has
effectively been used in Ref.~\cite{Khanbekyan} for analyzing the
quantum-state extraction from a cavity in the presence of unwanted
losses.  It is useful when the input field is in the vacuum state.
In this case the possibility of absorption or scattering of input
photons plays no role.

\subsection{Noisy coupling mirror}
\label{sec2A}

In the most general case one may use the input port of a cavity for
different purposes, e.g., for combining the intracavity and input
modes in an output mode. This possibility can be useful for the
quantum-state reconstruction as considered in Sec.~\ref{appl}. For a
correct description of such a process one should take into account
that input photons can be absorbed or scattered by the coupling
mirror before entering into the cavity.

\begin{figure}[ht!]
\includegraphics[width=
.95\linewidth, clip=]{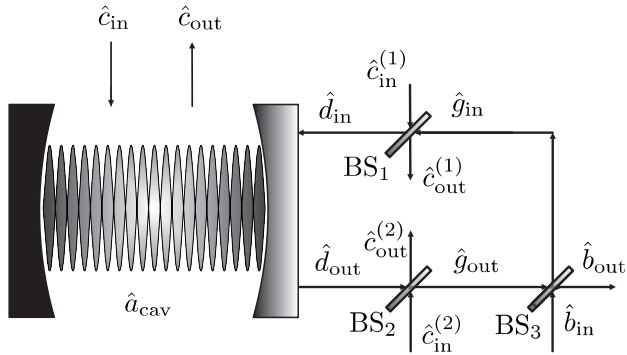} \caption{\label{fig5} Replacement
scheme for modeling the unwanted noise in a one-sided cavity. The
symmetrical SU(2)-type beam splitters $\mathrm{BS}_1$ and
$\mathrm{BS}_2$ model the unwanted noise in the coupling mirror, and
the asymmetrical U(2)-type beam splitter $\mathrm{BS}_3$ simulates
some feedback. }
\end{figure}

The corresponding type of unwanted noise can be included in the
theory in a systematic way by applying the concept of replacement
schemes as follows. Instead of considering the actual coupling
mirror, we consider an ideal semitransparent mirror that does not
give rise to unwanted noise and we model the unwanted noise by
inserting appropriately chosen beam splitters in the input and
output channels of the cavity, as sketched in Fig.~\ref{fig5}.
Clearly, the symmetric beam splitters BS$_1$ and BS$_2$,
respectively, are closely related to the unwanted losses that the
input and output fields suffer when passing through the coupling
mirror. Moreover it will turn out that a third beam splitter BS$_3$
is required, which simulates some feedback. This (asymmetric) beam
splitter is allowed to realize an U(2)-group transformation, thereby
introducing an additional phase shift (see Appendix~\ref{app1} and
Refs.~\cite{Vogel,Knoell2}). Including such a feedback into the
theory ensures one to describe all kinds of unwanted noise in
high-$Q$ cavities. The corresponding proof is given in
Sec.~\ref{sec3}.

Using Eqs.~(\ref{rs3}) and (\ref{rs4}) and the input-output
relations for each beam-splitter (Appendix \ref{app1}), we obtain
the extended quantum Langevin equation
\begin{align}
\label{rs25}
&\dot{\hat{a}}_\mathrm{cav}(t)=-\left[i\omega_\mathrm{cav}+
{\textstyle\frac{1}{2}}\Gamma\right]\hat{a}_\mathrm{cav}(t)\\
& \quad
 +\,\T^\mathrm{(c)}\hat{b}_\mathrm{in}\left(t\right)
+\A^\mathrm{(c)}_{(1)}\hat{c}^{(1)}_\mathrm{in}\left(t\right)
+\A^\mathrm{(c)}_{(2)}\hat{c}^{(2)}_\mathrm{in}\left(t\right)
+\A\hat{c}_\mathrm{in}\left(t\right)\nonumber
\end{align}
and the extended input-output relation
\begin{align}
\label{rs27}
\hat{b}_\mathrm{out}\left(t\right) = &\;
\T^\mathrm{(o)}\hat{a}_\mathrm{cav}\left(t\right)
+\R^\mathrm{(o)}\hat{b}_\mathrm{in}\left(t\right)
\nonumber\\ &
+\,\A^\mathrm{(o)}_{(1)}\hat{c}^{(1)}_\mathrm{in}\left(t\right)
+\A^\mathrm{(o)}_{(2)}\hat{c}^{(2)}_\mathrm{in}\left(t\right)
\end{align}
for a cavity in the presence of unwanted noise
(Appendix \ref{app2}).
Here,
\begin{equation}
\Gamma=\gamma\,\frac{1-\left|\R^{(3)}\right|^2\left|\T^{(1)}\right|^2
\left|\T^{(2)}\right|^2}
{\left|1-\R^{(3)\ast}\T^{(1)}\T^{(2)}\right|^2}
+\left|\A\right|^2
\label{rs35}
\end{equation}
is the cavity decay rate and
\begin{equation}
\omega_\mathrm{cav}
=\omega_{0}
- i\frac {\gamma} {2}
\frac{\R^{(3)\ast}\T^{(1)}\T^{(2)}-
\R^{(3)}\T^{(1)\ast}\T^{(2)\ast}}
{\left|1-\R^{(3)\ast}\T^{(1)}\T^{(2)}\right|^2}
\label{rs36}
\end{equation}
is the shifted frequency of the cavity mode.
The other $c$-number
coefficients
are defined as follows:
\begin{align}
&
\T^\mathrm{(c)}=\sqrt{\gamma}\,\frac{\T^{(1)}\T^{(3)\ast}}
{1-\R^{(3)\ast}\T^{(1)}\T^{(2)}}\,,\label{rs28}
\\[.5ex]&
\A^\mathrm{(c)}_{(1)}=\sqrt{\gamma}\,\frac{\R^{(1)}}
{1-\R^{(3)\ast}\T^{(1)}\T^{(2)}}\,,\label{rs29}
\\[.5ex]&
\A^\mathrm{(c)}_{(2)}=-\sqrt{\gamma}\,\frac{\T^{(1)}\R^{(2)}\R^{(3)\ast}}
{1-\R^{(3)\ast}\T^{(1)}\T^{(2)}}\,,\label{rs30}
\end{align}
\begin{align}
&
\T^\mathrm{(o)}=\sqrt{\gamma}\,e^{i\varphi^{(3)}}\frac{\T^{(2)}\T^{(3)}}
{1-\R^{(3)\ast}\T^{(1)}\T^{(2)}}\,,\label{rs31}
\\[.5ex]&
\R^\mathrm{(o)}=e^{i\varphi^{(3)}}\frac{\R^{(3)}-\T^{(1)}\T^{(2)}}
{1-\R^{(3)\ast}\T^{(1)}\T^{(2)}}\,,\label{rs32}
\\[.5ex]&
\A^\mathrm{(o)}_{(1)}=-e^{i\varphi^{(3)}}\frac{\T^{(2)}\R^{(1)}\T^{(3)}}
{1-\R^{(3)\ast}\T^{(1)}\T^{(2)}}\,,\label{rs33}
\\[.5ex]&
\A^\mathrm{(o)}_{(2)}=e^{i\varphi^{(3)}}\frac{\R^{(2)}\T^{(3)}}
{1-\R^{(3)\ast}\T^{(1)}\T^{(2)}}\,
,
\label{rs34}
\end{align}
where $\T^{(k)}$ and $\R^{(k)}$, respectively, are the transmission
and reflection coefficients of the $k$th beam splitter, and
$\varphi^{(3)}$ is a phase factor attributed to the third beam
splitter.

We see that the replacement scheme in Fig.~\ref{fig5} leads to a
description of the cavity in terms of the quantum Langevin equation
(\ref{rs25}) and input-output relation (\ref{rs27}) which are suited
to include unwanted noise in the theory. The corresponding
coefficients are expressed via the parameters of the component parts
of the replacement scheme---the cavity (with a coupling mirror that
is free of unwanted losses) and three beam splitters.  It is worth
noting that the results obtained are in agreement with those derived
on the basis of the QFT approach in Ref.~\cite{Khanbekyan2}.

\subsection
{Commutation relations}
\label{sec2B}

Clearly, the $c$-number coefficients in Eqs.~(\ref{rs25})
and (\ref{rs27}) are not independent of each other, since they
ensure, by construction, the validity of the
commutation relations
\begin{align}
&
[\hat{a}_\mathrm{cav}(t),
\hat{a}_\mathrm{cav}^\dag(t)]=1, \label{cp6}
\\&
\bigl[\hat{b}_\mathrm{out}(t_1),
\hat{b}_\mathrm{out}^\dag(t_2)
\bigr]=\delta(t_1-t_2)
\label{cp7}
\end{align}
Vice versa, if  the commutation relations (\ref{cp6}) and
(\ref{cp7}) are assumed to be valid, then
from the quantum Langevin equation
(\ref{rs25}) together with the in\-put-out\-put relation
(\ref{rs27})
and the commutation relations
\begin{align}
&
\bigl[\hat{b}_\mathrm{in}(t_1),
\hat{b}_\mathrm{in}^\dag(t_2)
\bigr]=\delta(t_1-t_2),
\label{cp2}
\\&
\bigl[\hat{c}_\mathrm{in}(t_1),
\hat{c}_\mathrm{in}^\dag(t_2)
\bigr]=\delta(t_1-t_2),
\label{cp3}
\\&
\bigl[\hat{c}_\mathrm{in}^{(1)}(t_1),
\hat{c}_\mathrm{in}^{(1)\dag}(t_2)
\bigr]=\delta(t_1-t_2),
\label{cp4}
\\&
\bigl[\hat{c}_\mathrm{in}^{(2)}(t_1),
\hat{c}_\mathrm{in}^{(2)\dag}(t_2)
\bigr]=\delta(t_1-t_2)
\label{cp5}
\end{align}
it necessarily follows that relations between
the mentioned coefficients must exist.
Note that mixed commutators vanish as a natural consequence of
the assumption that the cavity mode, the external modes, and
the dissipative systems responsible for the unwanted noise
are assumed to
refer to different degrees of freedom.

Inserting the
solution of the quantum Langevin
equation~(\ref{rs25})
\begin{align}
&\hat{a}_\mathrm{cav}(t) =
\hat{a}_\mathrm{cav}(0)e^{-(i\omega_\mathrm{cav}+\Gamma/2)t}
\nonumber\\&\hspace{2.5ex}
+\displaystyle{\int_{0}^{t}}\D t^{\prime}
e^{-(i\omega_\mathrm{cav}+\Gamma/2)(t-t^{\prime})}
\bigl[
\T^\mathrm{(c)}
\hat{b}_\mathrm{in}(t^{\prime})
\nonumber\\&\hspace{7.5ex}
+\A^\mathrm{(c)}_{(1)}\hat{c}^{(1)}_\mathrm{in}(t^{\prime})
+\A^\mathrm{(c)}_{(2)}\hat{c}^{(2)}_\mathrm{in}(t^{\prime})
+\A\hat{c}_\mathrm{in}(t^{\prime})\bigr]
\label{cp8}
\end{align}
in the left-hand side of Eq.~(\ref{cp6}),
assuming that
\begin{equation}
\left[\hat{a}_\mathrm{cav}(0),
\hat{a}_\mathrm{cav}^\dag(0) \right]=1
,
\label{cp1}
\end{equation}
and taking into account Eqs.~(\ref{cp2}--\ref{cp5}),
we find that Eq.~(\ref{cp6})
holds true only if the
condition
\begin{equation}
\Gamma=\bigl|\A\bigr|^2+\bigl|\A^\mathrm{(c)}_{(1)}\bigr|^2+
\bigl|\A^\mathrm{(c)}_{(2)}\bigr|^2+\bigl|\T^\mathrm{(c)}\bigr|^2
\label{cp9}
\end{equation}
is satisfied.
Similarly, inserting Eq.~(\ref{rs27}), together with
$\hat{a}_\mathrm{cav}(t)$ from Eq.~(\ref{cp8}), in the left-hand side
of Eq.~(\ref{cp7}),
we can easily see that Eq.~(\ref{cp7}) holds true if the conditions
\begin{equation}
\bigl|\R^\mathrm{(o)}\bigr|^2+\bigl|\A^\mathrm{(o)}_{(1)}\bigr|^2+
\bigl|\A^\mathrm{(o)}_{(2)}\bigr|^2=1 \label{cp10}
\end{equation}
and
\begin{equation}
\T^\mathrm{(o)}+\T^{\mathrm{(c)}\ast}\R^\mathrm{(o)}
+\A^{\mathrm{(c)}\ast}_{(1)}\A^\mathrm{(o)}_{(1)}
+\A^{\mathrm{(c)}\ast}_{(2)}\A^\mathrm{(o)}_{(2)}=0 \label{cp11}
\end{equation}
are satisfied.
Needless to say that
substituting Eqs.~(\ref{rs35}) and (\ref{rs28})--(\ref{rs34})
into
Eqs.~(\ref{cp9})--(\ref{cp11}) and utilizing
Eq.~(\ref{appeq0})
yields identities.

\section{Consistency and completeness}
\label{sec3}

There exist some other approaches to the problem of unwanted noise
in cavities, which may lead to quantum Langevin equations and
input-output relations different from Eqs.~(\ref{rs25}) and
(\ref{rs27}); see, e.g., Ref.~\cite{Viviescas}. Hence the question
of equivalence and completeness of different types of quantum
Langevin equations and the input-output relations associated with
them arises. Answering the question is no trivial task, and, in
fact, some approaches describe cavities which do not describe all
the typical situations.

Quite general, the quantum Langevin equation and the input-output
relation can be written in the form
\begin{equation}
\label{rep5}
\dot{\hat{a}}_\mathrm{cav}=-\left[i\omega_\mathrm{cav}+
{\textstyle \frac{1}{2} }
\Gamma\right]\hat{a}_\mathrm{cav}
 +\,\T^\mathrm{(c)}\hat{b}_\mathrm{in}\left(t\right)
+\hat{C}^{(c)}\left(t\right),
\end{equation}
\begin{equation}
\label{rep6}
\hat{b}_\mathrm{out}\left(t\right)=
\T^\mathrm{(o)}\hat{a}_\mathrm{cav}\left(t\right)
+\R^\mathrm{(o)}\hat{b}_\mathrm{in}\left(t\right)
+\hat{C}^{(o)}\left(t\right).
\end{equation}
where the operators $\hat{C}^{(c)}\!\left(t\right)$ and $\hat{C}^{(o)}\!\left(t\right)$
should obey the
commutation relations
\begin{align}
\label{rep6-1}
&\bigl[\hat{C}^{(c)}(t_1),\hat{C}^{(c)\dag}(t_2)\bigr]=
\bigl|\A^{(c)}\bigr|^2\delta(t_1-t_2),
\\
\label{rep6-2}
&\bigl[\hat{C}^{(o)}(t_1),\hat{C}^{(o)\dag}(t_2)\bigr]=
\bigl|\A^{(o)}\bigr|^2\delta(t_1-t_2),
\\
\label{rep6-3}
&\bigl[\hat{C}^{(c)}(t_1),\hat{C}^{(o)\dag}(t_2)\bigr]=
\left|\A^{(c)}\right|\left|\A^{(o)}\right|e^{i\kappa}\cos\zeta
\,\delta(t_1-t_2).
\end{align}
Here $\left|\A^{(c)}\right|$, $\left|\A^{(o)}\right|$, and
$e^{i\kappa}\cos\zeta$ are coefficients that along with $\T^{(c)}$,
$\T^{(o)}$, $\R^{(o)}$, and $\Gamma$ satisfy the constraints
\begin{equation}
\Gamma=\bigl|{\A}^\mathrm{(c)}\bigr|^2+
\bigl|\T^\mathrm{(c)}\bigr|^2,
\label{rep7}
\end{equation}
\begin{equation}
\bigl|\R^\mathrm{(o)}\bigr|^2+\bigl|\A^\mathrm{(o)}\bigr|^2=1, \label{rep8}
\end{equation}
\begin{equation}
\T^\mathrm{(o)}+\T^{\mathrm{(c)}\ast}\R^\mathrm{(o)}+
\left|\A^{(c)}\right|\left|\A^{(o)}\right|e^{i\kappa}\cos\zeta=0,\label{rep9}
\end{equation}
which follow, in a similar way as outlined for the scheme in
Sec.~\ref{sec2B}, from the requirement of preserving the commutation
rules.

The
constraints
(\ref{rep7})--(\ref{rep9})
[or
(\ref{cp9})--(\ref{cp11})
in the case of the scheme in Fig.~\ref{fig5}]
mean that the $c$-number coefficients
in Eqs.~(\ref{rep5}-\ref{rep6-3})
cannot be chosen freely, but
can take values
only
on a certain
manifold. In this context, Eqs.~(\ref{rs35})--(\ref{rs34}) can be
considered as an example of
a
parametrization of this manifold,
where
the number
of parameters describing the component part of
the replacement scheme in Fig.~\ref{fig5} exactly
coincides with the dimensionality of the manifold.

However, one may also think of parameterizations that do not cover
the whole manifold. In this case the corresponding replacement
scheme---referred to as a degenerate scheme---does not describe all
possible cavities. In order to test as to whether a given
parametrization is associated with a degenerate scheme, one can
apply an appropriate theorem of differential geometry
\cite{Dubrovin}. For this purpose one should first present
Eqs.~(\ref{rs35})--(\ref{rs34}) in the form of real functions of
real arguments. Next, one should find the rank of the matrix
constructed from the first derivatives of these functions and
compare it with the dimensionality of the manifold. For the
replacement scheme in Fig.~\ref{fig5} this has been checked using
\texttt{MATHEMATICA}.  As expected, it has turned out that the
scheme is nondegenerate. Hence the scheme leads to a complete and
consistent description of a (one-sided) cavity with unwanted noise.

Another possibility to express the $c$-number coefficients in terms
of independent parameters follows from Eqs.~(\ref{rep7}-\ref{rep9}).
One can simply consider the coefficients $\T^{(c)}$, $\T^{(o)}$,
$\R^{(o)}$, $\Gamma$ and $\omega_\mathrm{cav}$ as independent
parameters. The coefficients $\left|\A^{(c)}\right|$,
$\left|\A^{(o)}\right|$ and $e^{i\kappa}\cos\zeta$, describing the
unwanted noise, can be expressed in terms of them. The values of
such independent parameters have to belong to the manifold defined
by Eqs.~(\ref{rep7}-\ref{rep9}).

It is worth noting that the operators $\hat{C}^{(c)}(t)$ and
$\hat{C}^{(o)}(t)$ in Eqs.~(\ref{rep5}) and (\ref{rep6}) can be
expanded as
\begin{align}
&\hat{C}^{(c)}(t)=\sum_k
{\A}^\mathrm{(c)}_{(k)}\hat{{c}}^{(k)}_\mathrm{in}(t),
\label{rep3}\\
&\hat{C}^{(o)}(t)
=\sum_k
{\A}^\mathrm{(o)}_{(k)}\hat{{c}}^{(k)}_\mathrm{in}(t).
\label{rep4}
\end{align}
This implies that different representations (and corresponding replacement
schemes) of the operators of the unwanted noise can be obtained. The quantum
Langevin equation and the input-output relation in the form of Eqs.~(\ref{rs25})
and (\ref{rs27}) are an example of such a representation, where the
operators of the unwanted noise are expanded in a three-dimensional space.
However, it is clear that two operators $\hat{C}^{(c)}(t_1)$ and
$\hat{C}^{(o)}(t_1)$ can be expanded in two-dimensional basis. Hence, for the
complete characterization of a cavity, one can use only two independent sources of unwanted noise.

\begin{figure}[ht!]
\includegraphics[width=.95\linewidth,clip=]{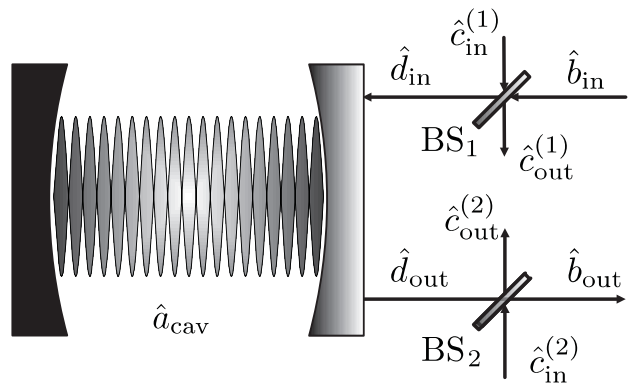}
\caption{\label{fig9} An example of a degenerate replacement
scheme.}
\end{figure}

As already mentioned, there exist schemes which do not describe all
the physically possible lossy cavities and, in fact, describe
special cases of lossy cavities. In other words, the corresponding
parameters do not cover the whole manifold of the values of the
coefficients in Eqs.~(\ref{rep5}) and (\ref{rep6}) which are, in
principle, possible and hence, the schemes can be considered as
being degenerate. An example of such scheme is the replacement
scheme in Fig.~\ref{fig9}. The corresponding quantum Langevin
equation and the input-output relation, which are special cases of
Eqs.~(\ref{rep5}) and (\ref{rep6}), read
\begin{align}
\label{rep34}
&\dot{\hat{a}}_\mathrm{cav}= -\left[i\omega_\mathrm{cav}+
{\textstyle \frac{1}{2}}
\Gamma\right]\hat{a}_\mathrm{cav}
 + \T^\mathrm{(c)}\hat{b}_\mathrm{in}(t)
+{\A}^\mathrm{(c)}_{(1)}\hat{{c}}^{(1)}_\mathrm{in}(t)
,
\\
\label{rep35}
&\hat{b}_\mathrm{out}(t)=
\T^\mathrm{(o)}\hat{a}_\mathrm{cav}(t)
+\R^\mathrm{(o)}\hat{b}_\mathrm{in}(t)
\nonumber\\ &\hspace{9ex}
+  {\A}^\mathrm{(o)}_{(1)}\hat{c}^{(1)}_\mathrm{in}(t)
+{\A}^\mathrm{(o)}_{(2)}\hat{c}^{(2)}_\mathrm{in}(t)
\end{align}
The
parametrization
can easily be obtained from the parametrization
(\ref{rs35})--(\ref{rs34}) by setting therein $\T^{(3)}$
$\!=$ $\!1$, $\R^{(3)}$ $\!=$ $\!0$ and $\A$
$\!=$ $\!0$.

It is not difficult to prove that for this scheme, along with
Eqs.~(\ref{rep7})--(\ref{rep9}), the additional constraint
\begin{equation}
\frac{\T^\mathrm{(o)}\T^\mathrm{(c)}}{\Gamma}+\R^\mathrm{(o)}=0
\label{rep45}
\end{equation}
is satisfied. This relation does not follow from the requirement of
preserving the commutation rules (\ref{cp6}) and (\ref{cp7}).
Clearly, the rank of the matrix constructed from the first
derivatives of the real functions corresponding to the
parametrization is not equal to the number of the independent
coefficients in Eqs.~(\ref{rep34}) and (\ref{rep35})---a sign that
the scheme is indeed degenerate.

It is worth noting that the physics behind this degenerate scheme is
closely related to that of a cavity without unwanted noise. This
becomes clear from the following argument. The loss channels modeled
by the two beam splitters may equivalently be interpreted as the
losses that the input (output) field suffers from before entering
(after leaving) the cavity. A consequence of this fact is that the
losses modeled in this way cannot affect the decay rate of the
intracavity mode. Thus the unwanted losses do not affect the
dynamics of the intracavity mode.

\section{Noise-induced mode coupling}
\label{sec5}

In the generation and processing of nonclassical radiation one is
commonly interested in a reduction of unwanted noise, because it
gives rise to quantum decoherence, in general. However, if the input
port of a cavity is used, the presence of unwanted losses does not
only change the properties of the intracavity mode and the outgoing
field. In this case, a new possibility for combining the intracavity
mode and an input mode in a nonmonochromatic output mode appears---a
surprising property, which does not exist for cavities without
unwanted noise channels. Moreover, such an effect cannot be properly
described by a degenerate cavity model such as that given in
Fig~\ref{fig9}.

After sufficiently long time, $t\gg 1/\Gamma$, the internal field of
an ideal cavity (i.e., a cavity without unwanted noise) is
completely transferred into the nonmonochromatic cavity-associated
output mode (CAOM). Since the efficiency of this process is equal to
one, an input signal cannot be reflected into this mode. Therefore,
in order to combine the intracavity mode and an input mode in an
output mode, one must decrease this efficiency. Realistic cavities
are always characterized by some unwanted losses, such as absorption
and scattering. Hence, the efficiency of intracavity mode escaping
from such a cavity is less than unity \cite{Khanbekyan} and an input
mode can be reflected, in principle, into the CAOM. Therefore,
combining the intracavity mode with an input mode in the output mode
becomes possible.

Assuming that the quantum state of intracavity mode is generated at
the zero point of time, the solution of the quantum Langevin equation
(\ref{rep5}) can be written as
\setlength\arraycolsep{0pt}
\begin{multline}
\label{SolQLE}
\hat{a}_\mathrm{cav}\left(t_1\right)=
\hat{a}_\mathrm{cav}\left(0\right)\T^{\mathrm{(o)}-1}F^{\ast}\left(t_1\right)
\\
+\T^\mathrm{(c)}\T^{\mathrm{(o)}-1}\int_{0}^{t_1}\D t_2\,
\xi^{\ast}\left(t_1,t_2\right)
\hat{b}_\mathrm{in}\left(t_2\right)
+\hat{\tilde{C}}\left(t_1\right),
\end{multline}
where
\begin{equation}
F^\ast\left(t_1\right)=
\T^\mathrm{(o)}e^{-\left(i\omega_\mathrm{cav}+\frac{\Gamma}{2}\right)t_1}
\Theta\left(t_1\right), \label{FTime}
\end{equation}
\begin{equation}
\xi^\ast\left(t_1,t_2\right)= \T^\mathrm{(o)}
e^{-\left(i\omega_\mathrm{cav}+\frac{\Gamma}{2}\right)\left(t_1-t_2\right)}
\Theta\left(t_1\right)\Theta\left(t_1-t_2\right),\label{XiTime}
\end{equation}
$\Theta\left(t_1\right)$ is a unit step function and
$\hat{\tilde{C}}\left(t_1\right)$ is a linear integral expression
containing the operators of unwanted noise. Since we assume that the
unwanted-noise systems are in the vacuum state, the explicit form of
$\hat{\tilde{C}}\left(t_1\right)$ plays no role for our further
consideration. Substituting Eq.~(\ref{SolQLE}) into the input-output
relation (\ref{rep6}), one obtains the relation
\begin{multline}
\hat{b}_\mathrm{out}\left(t_1\right)=\label{IORTime}
\hat{a}_\mathrm{cav}(0)F^\ast\left(t_1\right)
\\
+\int_{-\infty}^{+\infty}\!\D
t_2\, G^\ast\left(t_1,t_2\right)\hat{b}_\mathrm{in}\left(t_2\right)
+\hat{C}\left(t_1\right).
\end{multline}
Hence, the output-mode operator is expressed by the input-mode
operator, the intracavity mode operator at the initial time and the
operators of unwanted noise. Here
 \begin{equation}
G^\ast\left(t_1,t_2\right)=\T^\mathrm{(c)}
\xi^\ast\left(t_1,t_2\right)+\R^\mathrm{(o)}\delta
\left(t_1-t_2\right), \label{GTime}
\end{equation}
and the operator $\hat{C}\left(t_1\right)$ is again a linear
integral expression containing the operators of unwanted noise,
whose explicit form is not needed for the further considerations.
The first term in Eq.~(\ref{IORTime}) describes the extraction of
the intracavity mode into the CAOM. The second term describes the
reflection of the input field, where $G^{\ast}\left(t_1,t_2\right)$
is the integral kernel of the corresponding mode transformation. It
is worth noting that non-Hermitian properties of this integral
transformation lead to changing (decreasing) the norm of the
reflected pulse compared with the input one. This corresponds to the
partial absorption/scattering during reflection at the cavity.

For our purposes it is convenient to use another (equivalent)
representation of Eq.~(\ref{IORTime}). Let
$\left\{U_n^\mathrm{in}\left(t_1\right), n=0,\ldots,+\infty\right\}$
and $\left\{U_n^\mathrm{out}\left(t_1\right),
n=0,\ldots,+\infty\right\}$  be two different complete sets of
orthogonal functions associated with the input and output modes
respectively, i.e.,
\begin{equation}
\hat{b}_\mathrm{in(out)}\!\left(t_1\right)=\sum\limits_{n=0}^{+\infty}
U_n^\mathrm{in(out)}\left(t_1\right)\hat{b}_\mathrm{in(out);n}\,
,\label{BDiscrete1}
\end{equation}
\begin{equation}
\hat{b}_\mathrm{in(out); n}=\int_{-\infty}^{+\infty} \D t_1
U_n^{\mathrm{in(out)}\ast}\!\left(t_1\right)
\hat{b}_\mathrm{in(out)}\!\left(t_1\right).\label{BDiscrete2}
\end{equation}
Here $\hat{b}_{\mathrm{in(out)}; n}$ is the annihilation operator of
an input (output) photon in the nonmonochromatic mode corresponding
to the function $U_n^{\mathrm{in(out)}}\!\left(t_1\right)$. We
choose the function $U_0^\mathrm{out}\!\left(t_1\right)$ in the form
of the CAOM,
\begin{equation}
U_0^\mathrm{out}\!\left(t_1\right)=\frac{F^\ast\left(t_1\right)}
{\sqrt{\eta_\mathrm{ext}}},\label{U0Out}
\end{equation}
where
\begin{equation}
\eta_\mathrm{ext}=\int_{-\infty}^{+\infty} \D t_1
\left|F\left(t_1\right)\right|^2
=\frac{\left|\T^{(o)}\right|^2}{\Gamma}\label{EtaExt}
\end{equation}
can be interpreted as the efficiency of the intracavity-field
extraction into the CAOM \cite{Khanbekyan}. The function
$U_0^\mathrm{in}\!\left(t_1\right)$, defined by using the integral
kernel $G\left(t_2,t_1\right)$ as
\begin{align}
U_0^\mathrm{in}\!\left(t_1\right)
&=\frac{1}{\sqrt{\underline{\eta}_\mathrm{ref}}}
\int_{-\infty}^{+\infty}\D t_2\, G\!\left(t_2,t_1\right)U_0^\mathrm{out}\!
\left(t_2\right)
\nonumber\\[1ex]
&=U_0^\mathrm{out}\!
\left(t_1\right)e^{-i\varphi},\label{U0In}
\end{align}
corresponds to the nonmonochromatic matched input mode (MIM), which
only makes a contribution, among the other orthogonal input modes of
this set, into the CAOM under reflection at the cavity. Here
\begin{equation}
\underline\eta_\mathrm{\, ref}=
\left|\frac{\T^\mathrm{(o)}\T^\mathrm{(c)}}{\Gamma}
+\R^\mathrm{(o)}\right|^2\label{EtaRefCAOM}
\end{equation}
is the efficiency of the MIM reflection into the CAOM, which can be
found through the condition of normalization for the function
$U_0^\mathrm{in}\!\left(t_1\right)$, Eq.~(\ref{U0In}). The phase
$\varphi$ is defined as
\begin{equation}
\varphi=\arg\left[\frac{\T^\mathrm{(o)}\T^\mathrm{(c)}}{\Gamma}+\R^\mathrm{(o)}\right].\label{Phase1}
\end{equation}

\begin{figure}%[ht!]
\includegraphics[width=0.45\textwidth, clip=]{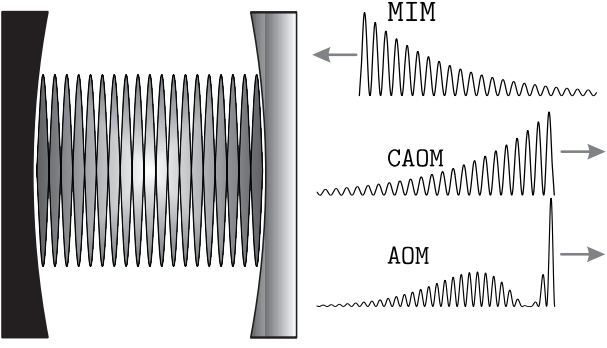}
\caption{\label{fig11} The mode structure of the external field:
cavity-associated output mode (CAOM), additional output mode (AOM),
and matched input mode (MIM).}
\end{figure}

Along with the CAOM, the MIM is reflected into another
nonmonochromatic output mode as well, see Fig.~\ref{fig11}. This
additional output mode (AOM) results in noise effects when one measures some
properties of the quantum state of the CAOM. To analyze it, we
need the total response of the cavity on the MIM, that can be
obtained by using the integral kernel
$G^{\ast}\left(t_1,t_2\right)$ as
\setlength\arraycolsep{0pt}
\begin{eqnarray}
U^\mathrm{out}\!\left(t_1\right)&&=\int_{-\infty}^{+\infty}\D t_2\, G^{\ast}\!\left(t_1,t_2\right)U_0^\mathrm{in}\!\left(t_2\right)\label{MIMResponse}\\
&&=\sqrt{\Gamma}\!\left(\T^\mathrm{(c)}\T^\mathrm{(o)}t_1+\R^\mathrm{(o)}\right)\!
\nonumber\\ &&\times
e^{-\left(i\omega_\mathrm{cav}+\frac{\Gamma}{2}\right)t_1+i\left(\arg\T^\mathrm{(o)}-\varphi\right)}\Theta\left(t_1\right)
 .\nonumber
\end{eqnarray}

Since the total reflected pulse $U^\mathrm{out}\!\left(t_1\right)$
is a superposition of the CAOM with the AOM, i.e.,
\begin{equation}
U^\mathrm{out}\!\left(t_1\right)=\sqrt{\underline\eta_\mathrm{\,
ref}} U_0^\mathrm{out}\!\left(t_1\right)+
\sqrt{\overline\eta_\mathrm{ref}}
U_1^\mathrm{out}\!\left(t_1\right),\label{MIMResponseExp}
\end{equation}
the form of the AOM, denoted as
$U_1^\mathrm{out}\!\left(t_1\right)$, can be found as
\begin{align}
U_1^\mathrm{out}\!\left(t_1\right)
&=\frac{1}{\sqrt{\overline\eta_\mathrm{ref}}}
\left[U^\mathrm{out}\!\left(t_1\right)-\sqrt{\underline\eta_\mathrm{\, ref}}\,
U_0^\mathrm{out}\!\left(t_1\right)\right]\label{AdOM}\\
&=\sqrt{\Gamma}e^{i\chi}\left(\Gamma
t_1-1\right)e^{-\left(i\omega_\mathrm{cav}+\frac{\Gamma}{2}\right)t_1}
\Theta\left(t_1\right).\nonumber
\end{align}
Here
\begin{equation}
\chi=\arg\frac{\T^\mathrm{(o)}\T^\mathrm{(c)}}{\Gamma}+\arg\T^\mathrm{(o)}-\varphi\label{Phase2}
\end{equation}
and
\begin{equation}
\overline\eta_\mathrm{ref}
=\frac{\left|\T^\mathrm{(o)}\right|^2\left|\T^\mathrm{(c)}\right|^2}{\Gamma^2}\label{EtaRefAdOM}
\end{equation}
is the efficiency of the reflection
of the MIM into the AOM, which is found via the normalization of the function
$U_0^\mathrm{out}\!\left(t_1\right)$.

One can check by direct calculations that in the new representation
Eq.~(\ref{IORTime}) reads
\begin{equation}
\hat{b}_{\mathrm{out}; 0}=
\sqrt{\eta_\mathrm{ext}}\,\hat{a}_\mathrm{cav}\!\left(0\right)+
\sqrt{\underline\eta_\mathrm{\, ref}}\,\hat{b}_{\mathrm{in}; 0}
+\hat{C}_{0},\label{IORDiscrete0}
\end{equation}
\begin{equation}
\hat{b}_{\mathrm{out;} 1}=
\sqrt{\overline{\eta}_\mathrm{ref}}\,\hat{b}_{\mathrm{in}; 0}+
\sum\limits_{m=1}^{+\infty}\! G^\ast_{m,1}\,\hat{b}_\mathrm{in; m}+\hat{C}_{1},\label{IORDiscrete1}
\end{equation}
\begin{equation}
\hat{b}_{\mathrm{out}; n}= \sum\limits_{m=1}^{+\infty}\!
G^\ast_{m,n}\,\hat{b}_{\mathrm{in}; m}+\hat{C}_{n}\, \textrm{for
$n=2,3,\ldots$}, \label{IORDiscrete2}
\end{equation}
where
\begin{equation}
G^\ast_{m,n}= \int_{-\infty}^{+\infty}\D t_1 \D t_2
U_{n}^{\mathrm{out}\ast}\!\left(t_1\right)G^\ast\left(t_1,t_2\right)
U_{m}^\mathrm{in}\left(t_2\right),\label{GDiscrete}
\end{equation}
\begin{equation}
\hat{C}_{n}=\int_{-\infty}^{+\infty}\D t_1
U_{n}^{\mathrm{out}\ast}\! \left(t_1\right)
\hat{C}\left(t_1\right).\label{CDiscrete}
\end{equation}
The first term of Eq.~(\ref{IORDiscrete0}) describes the
intracavity-field extraction into the CAOM with the efficiency
$\eta_\mathrm{ext}$ \cite{Khanbekyan}. This mode corresponds to the
function $U_0^\mathrm{out}\!\left(t_1\right)$. The second term of
Eq.~(\ref{IORDiscrete0}) demonstrates the possibility to combine the
MIM and the intracavity mode in the CAOM with the efficiency
$\underline\eta_\mathrm{\, ref}$ given by Eq.~(\ref{EtaRefCAOM}). As
it follows from Eqs.~(\ref{IORDiscrete1}) and (\ref{IORDiscrete2}),
the field extracted from the cavity does not give a contribution to
other nonmonochromatic output modes. Moreover, according to
Eq.~(\ref{IORDiscrete0}), only the MIM described by the function
$U_0^\mathrm{in}\!\left(t_1\right)$ contributes into the CAOM via
reflection at the cavity. It is worth noting that the MIM can be
easily prepared in an experiment since it has the form of a pulse
extracted from another cavity of the same type.

The frequency representation of the CAOM and the AOM have a very
similar form. Their Fourier images, denoted as
$U_0^\mathrm{out}\!\left(\omega\right)$ and
$U_1^\mathrm{out}\!\left(\omega\right)$ respectively, have equal
absolute values, i.e.,
\begin{equation}
\left|U_0^\mathrm{out}\!\left(\omega\right)\right|^2
=\left|U_1^\mathrm{out}\!\left(\omega\right)\right|^2
=\frac{\Gamma}{2\pi\left[\left(\omega-\omega_\mathrm{cav}\right)^2
+\frac{\Gamma^2}{4}\right]}.\label{Spectrum}
\end{equation}
Hence, these two orthogonal modes are irradiated in the same frequency domain. They
differ only in the phases.

For the cavity associated with the degenerate replacement scheme in
Fig.~\ref{fig9}, the efficiency of the reflection of the MIM into
the CAOM, see Eq.~(\ref{EtaRefCAOM}), is zero due the constraint
(\ref{rep45}). Therefore, the incomplete model does not describe the
possibility of the input and intracavity mode matching. Cavities
without channels of unwanted noise will not give rise to the mode
matching as well.

\section{Application to quantum-state reconstruction}
\label{appl}

The considered mode-coupling effect can be used for unbalanced
\cite{Wallentowitz} and cascaded \cite{Kis} homodyning of the
intracavity mode. Presently known methods for the reconstruction of
the quantum state of the intracavity mode are based either on an
interaction between atoms and intracavity field
\cite{AtomReconstruction} or on the balanced homodyning of the
extracted field \cite{Santos}. Including in the model unwanted
noise, which exists for all realistic cavities, allows one to
formulate another way for the quantum-state reconstruction of the
intracavity mode.

The proposed method has two major advantages. First, unbalanced
homodyning allows one to perform a local reconstruction of the
quantum state of the intracavity mode. That is, in contrast to
balanced homodyning, in unbalanced homodyning it is not required to
perform complicated integral transformations of measured data.
Second, one can directly use properties of the cavity to combine the
signal field with the local oscillator, which allows to avoid losses
associated with the additional beam splitter. Both features are
important in the considered case, because the quantum-state
extraction is typically characterized by a small efficiency
\cite{Hood} that gives additional difficulties in the numerical
evaluation of the measured data.

\subsection
{Unbalanced homodyning} \label{sec5B}

Let us assume that a quantum state of radiation has been generated
inside the cavity. The local-oscillator field with the coherent
amplitude $\beta$ is prepared in the form of the MIM, e.g., it could
be extracted from another cavity. The further calculations are
similar to those in Ref.~\cite{Wallentowitz}. The difference is that
the influence of the AOM must be taken into account. The
photodetector counts the photon number of the total outgoing field
\begin{equation}
\hat{n}_\mathrm{out}=\hat{b}_{\mathrm{out};
0}^\dag\,\hat{b}_{\mathrm{out}; 0} +\hat{b}_{\mathrm{out};
1}^\dag\,\hat{b}_{\mathrm{out}; 1}+\ldots , \label{nOut}
\end{equation}
The probability of recording $n$ counts reads
\begin{equation}
p_{n}=\left\langle:\frac{\left(\eta_\mathrm{c}\hat{n}_\mathrm{out}\right)^n}
{n!} \exp\left(-\eta_\mathrm{c}\hat{n}_\mathrm{out}\right)
:\right\rangle,\label{counts}
\end{equation}
where $::$ means normally ordering and $\eta_\mathrm{c}$ denotes the
counting efficiency. The $s$-parametrized phase space distribution
$P_\mathrm{cav}\!\left(\alpha;s\right)$ of the intracavity mode is
expressed in terms of the Glauber-Sudarshan $P$ distribution
$P_\mathrm{cav}\!\left(\alpha\right)$ in the form \cite{Cahill}
\begin{multline}
P_\mathrm{cav}\!\left(\alpha,s\right)=
\\
\frac{2}{\pi\left(1-s\right)}
\int\D^2\alpha^{\prime}
\exp\left({-\frac{2\left|\alpha-\alpha^{\prime}\right|^2}{1-s}}\right)P_\mathrm{cav}
\!\left(\alpha^{\prime}\right),
\end{multline}
which can be rewritten as
\begin{equation}
P_\mathrm{cav}\!\left(\alpha,s\right)=\frac{2}{\pi\left(1-s\right)}
\left\langle:\exp\left(-\frac{2}{1-s}\,\hat{n}_\mathrm{cav}\!\left(\alpha\right)\right):
\right\rangle, \label{distribution}
\end{equation}
where
\begin{equation}
\hat{n}_\mathrm{cav}\!\left(\alpha\right)=
\left(\hat{a}_\mathrm{cav}^{\dag}\!\left(0\right)-\alpha^{\ast}\right)
\left(\hat{a}_\mathrm{cav}\!\left(0\right)-\alpha\right)
\end{equation}
is the displaced photon-number operator of the intracavity mode.
Utilizing the input-output relations (\ref{IORDiscrete0}),
(\ref{IORDiscrete1}) and Eq.~(\ref{nOut}), and assuming that the MIM
is in a coherent state of amplitude $\beta$ and all other modes are
in the vacuum state, the operator
$\hat{n}_\mathrm{cav}\!\left(\alpha\right)$ in
Eq.~(\ref{distribution}) can be written in the form
\begin{equation}
\hat{n}_\mathrm{cav}\!\left(\alpha\right)=\frac{1}{\eta_\mathrm{ext}
}\,\hat{n}_\mathrm{out}-\epsilon\left|\alpha\right|^2,\label{nCav}
\end{equation}
where
\begin{equation}
\alpha=-\sqrt{\frac{\underline\eta_\mathrm{\,
ref}}{\eta_\mathrm{ext}}}\,\beta,\label{Amplitude}
\end{equation}
and the factor $\epsilon=\overline\eta_\mathrm{ref}/
\underline\eta_\mathrm{\, ref}$ can be rewritten with
the help of
Eqs.~(\ref{EtaRefCAOM}), (\ref{EtaRefAdOM}) as
\begin{equation}
\epsilon=\frac{1}{\left|1+ \frac{\R^\mathrm{(o)}\Gamma}
{\T^\mathrm{(o)}\T^\mathrm{(c)}} \right|^2}.\label{Epsilon}
\end{equation}
The second term in Eq.~(\ref{nCav}), caused by nonzero $\epsilon$,
describes the influence of the AOM.

This gives a possibility to rewrite Eq.~(\ref{distribution}) in the
form
\begin{multline}
P_\mathrm{cav}\!\left(\alpha,s\right)=
  \frac{2}{\pi\left(1-s\right)}
\exp\left(\frac{2}
{1-s}\epsilon\left|\alpha\right|^2\right)
\\[1ex]
\times
\left\langle:\exp\left(-\frac{2}{\left(1-s\right)\eta_\mathrm{ext}}
\,\hat{n}_\mathrm{out}\right): \right\rangle.
\label{Mdistribution}
\end{multline}
Similar to Ref.~\cite{Wallentowitz}, one can decompose the normally
ordered exponent into the factor
$\exp\left(-\eta_\mathrm{c}\hat{n}_\mathrm{out}\right)$ contained in
Eq.~(\ref{counts}) and a residual factor,
\begin{multline}
P_\mathrm{cav}\!\left(\alpha,s\right)=
\frac{2}{\pi\left(1-s\right)}
\exp\left(\frac{2}
{1-s}\epsilon\left|\alpha\right|^2\right)\\[1ex]
\times \left\langle:\exp\left(-\xi
\,\eta_\mathrm{c}\hat{n}_\mathrm{out}\right)
\exp\left(-\eta_\mathrm{c}\hat{n}_\mathrm{out}\right):
\right\rangle,
\label{MMdistribution}
\end{multline}
where
\begin{equation}
\xi=\frac{2- \eta(1-s)}{\eta(1-s)} \label{Xi}
\end{equation}
and $\eta$ is the overall efficiency of detection
\begin{equation}
\eta=\eta_\mathrm{ext}\eta_\mathrm{c}. \label{EtaOverall}
\end{equation}
Expanding the residual factor into a series, it is straightforward
to find the $s$-parametrized phase-space distribution
\begin{equation}
\begin{array}{ll}
P_\mathrm{cav}\!\left(\alpha;s\right)=\frac{2}{\pi\left(1-s\right)}\,&\exp\!\left(\frac{2}
{1-s}\epsilon\left|\alpha\right|^2\right)\\
&\times\sum\limits_{n=0}^{+\infty}\left(-\xi\right)^n
p_n\!\left(\alpha;\eta,
\epsilon\right),\end{array}\label{Reconstruction}
\end{equation}
where $p_n\!\left(\alpha;\eta, \epsilon\right)\equiv p_n$ is the
probability of recording $n$ counts given by Eq.~(\ref{counts}).

Thus, measuring the photocount statistics of the outgoing field
$p_n$, one can reconstruct the $s$-parametrized phase-space
distribution of the intracavity mode. It is worth noting that such a
reconstruction is impossible for cavities without channels of
unwanted noise and cavities whose channels of unwanted noise can be
modeled, for example, by a degenerate replacement scheme of the type
shown in Fig.~\ref{fig9}. From Eq.~(\ref{Amplitude}) it follows that
for such cavities $\underline\eta_\mathrm{\, ref}=0$, hence the
reconstruction is only possible for $\alpha=0$, i.e., for the origin
of the phase space. In contrast, if the unwanted noise sources are
described properly, the complete information about a quantum state
of the intracavity mode can be obtained.

Unlike the case considered in Ref.~\cite{Wallentowitz}, the
reliability of the method depends not only on the value of the
parameter $\xi$, but also on the parameter $\epsilon$, cf.
Eqs~(\ref{Xi}) and (\ref{Epsilon}). For the best convergence of the
series in Eq.~(\ref{Reconstruction}), both these parameters should
be less than unity. Nevertheless, it is impossible to satisfy these
two conditions simultaneously in realistic situations.

\begin{figure}[ht!]
\includegraphics[width=0.45\textwidth, clip=]{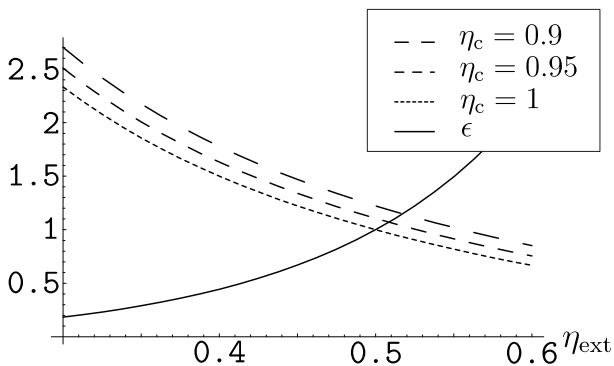}
\caption{\label{plots} The dependences of $\epsilon$ and $\xi$ on
$\eta_\mathrm{ext}$ for the reconstruction of the Husimi-Kano $Q$
function, $s=-1$, where $\xi$ is given for three different
efficiencies of photocounting $\eta_\mathrm{c}$.}
\end{figure}

\begin{figure}[ht!]
\textrm{(a)}\includegraphics[width=0.45\textwidth, clip=]{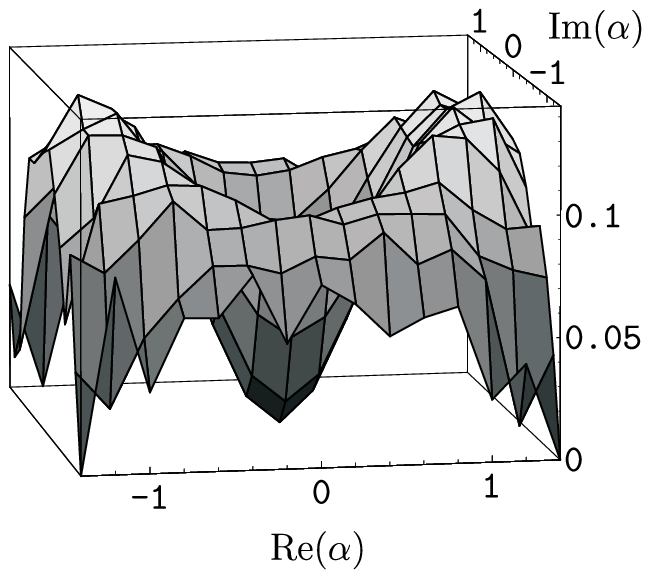}\\
\textrm{(b)}\includegraphics[width=0.45\textwidth,
clip=]{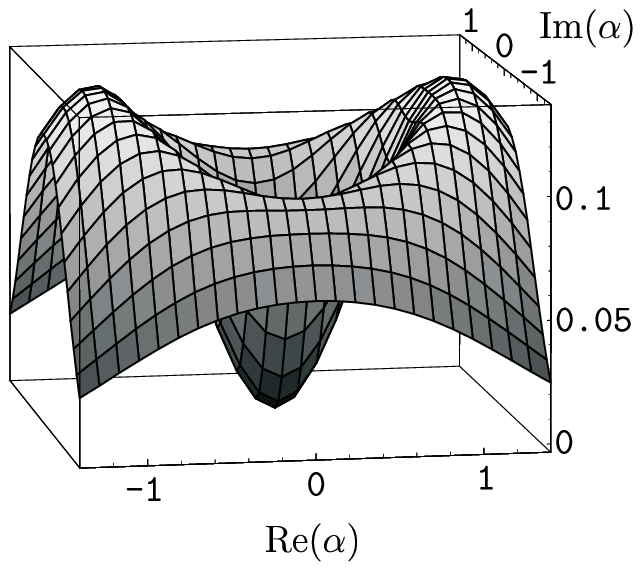} \caption{\label{Data} Reconstruction of the
Husimi-Kano $Q$ function for the odd superposition of the coherent
states
$\left|\delta_{-}\right\rangle=\mathcal{N}\big(\left|\delta\right\rangle-
\left|-\delta\right\rangle\big)$, $\delta=0.7$, by unbalanced
homodyning: \textrm{(a)} numerical simulation with $1.7\times 10^5$
sampling events for each point; \textrm{(b)} exact function.}
\end{figure}

To illustrate the method, we consider a cavity with zero absorption
coefficient $\A^\mathrm{(o)}$ of the input field, i.e., with
$\left|\R^\mathrm{(o)}\right|^2=1$. As it follows from the
constraints (\ref{rep7}-\ref{rep9}) the value of parameter
$\epsilon$ given by Eq.~(\ref{Epsilon}) can be written in the form
\begin{equation}
\epsilon=\left(\frac{\eta_\mathrm{ext}}{1-\eta_\mathrm{ext}}\right)^2.\label{Epsilon2}
\end{equation}
This is a rising function of $\eta_\mathrm{ext}$, in contrast to the
dependence on $\eta_\mathrm{ext}$ of $\xi$, see Fig.~\ref{plots}.
The intersection of these curves can be considered as the optimum
value of $\eta_\mathrm{ext}$. Even for an ideal detector with
$\eta_\mathrm{c}=1$, this value is $\eta_\mathrm{ext}=0.5$ and
corresponding value of the parameters $\xi$ and $\epsilon$ is
$\xi=\epsilon=1$.
Let us consider a photodetector with $\eta_\mathrm{c}=0.95$. In this
case, the optimum value of the efficiency of quantum-state
extraction is $\eta_\mathrm{ext}=0.5085$ which corresponds to the
situation in Ref.~\cite{Hood}. The corresponding values of the
parameters $\xi$ and $\epsilon$ for this case are
$\xi=\epsilon=1.070$. In Fig.~\ref{Data} the result of a numerical
simulation of the reconstruction of the Husimi-Kano $Q$ function for
the odd superposition of coherent states is presented. Each point is
evaluated with \mbox{$1.7\times 10^5$} sampling events. This is much
more than the number of \mbox{$5\times 10^3$} events needed for such
an experiment in the case of usual unbalanced homodyning with the
same overall efficiency. In particular, for the phase-space
distributions far from the origin of the phase space one needs a
large number of sampling events. The method works well for small
values of $|\alpha|$, for which $\sim 10^4$ sampling events are
sufficient.

\subsection
{Cascaded homodyning}

\begin{figure}%[ht!]
\includegraphics[width=0.45\textwidth, clip=]{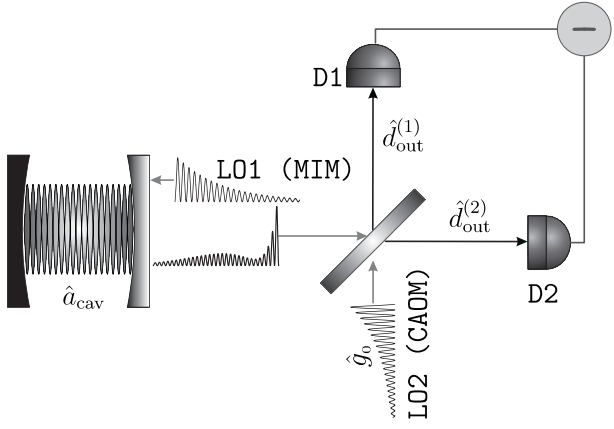}
\caption{\label{Cascaded} Cascaded homodyne detection of the
intracavity mode.}
\end{figure}

The efficiency of the scheme can be sufficiently improved by using
the related scheme of cascaded homodyning \cite{Kis}. In this scheme
the balanced homodyne detection is used for counting photons
\cite{Munroe} in the scheme of unbalanced homodyning, see
Fig.~\ref{Cascaded}. The local oscillator 1 ($\mathrm{LO1}$) is
prepared in the form of the MIM similar to the case of unbalanced
homodyning. The phase randomized local oscillator 2 ($\mathrm{LO2}$)
is prepared in the form of the CAOM and it can be derived from the
MIM, cf. Eq.~(\ref{U0In}). In this case the influence of the AOM
disappears completely. Hence the results of the work \cite{Kis} with
the overall efficiency given by Eq.~(\ref{EtaOverall}) can be
directly applied to this case.

Let us consider this scheme in more details. The photodetectors
$\mathrm{D1}$ and $\mathrm{D2}$ count the photon numbers in the
output ports of a $50:50$ beam splitter,
\begin{equation}
\hat{n}_\mathrm{out}^{(k)}=\hat{d}_{\mathrm{out};
0}^{\dag(k)}\,\hat{d}_{\mathrm{out}; 0}^{(k)}+\hat{d}_{\mathrm{out};
1}^{\dag (k)}\,\hat{d}_{\mathrm{out};1}^{(k)}+\ldots , \label{nOutCascade}
\end{equation}
where $k=1,2$ is the number of the detector. The operators
$\hat{d}_{\mathrm{out}; n}^{(k)}$ are connected with the mode
operators $\hat{b}_{\mathrm{out}; n}$ of the cavity output and the
operators $\hat{g}_\mathrm{n}$ describing the $\mathrm{LO2}$ via the
relations
\begin{align}
&\hat{d}_{\mathrm{out};
n}^{(1)}=\frac{1}{\sqrt{2}}\left(\hat{b}_{\mathrm{out}; n}
+\hat{g}_n\right),\\
&\hat{d}_{\mathrm{out};
n}^{(2)}=\frac{1}{\sqrt{2}}\left(-\hat{b}_{\mathrm{out}; n}
+\hat{g}_n\right).
\end{align}
The $\mathrm{LO2}$ corresponds to the MIM in a phase-randomized
coherent state, i.e.,
\begin{equation}
\hat{g}_\mathrm{0}\rightarrow re^{i\varphi},\\
\end{equation}
where $r$ is the amplitude and $\varphi$ is the random phase.
Whereas the AOM is superposed with the corresponding mode of
$\mathrm{LO2}$ being in the vacuum state
\begin{equation}
\hat{g}_\mathrm{1}\rightarrow 0.
\end{equation}
The difference of the photocounts $\hat{n}_\mathrm{out}^{(1)}$ and
$\hat{n}_\mathrm{out}^{(2)}$ can be written in the form
\begin{equation}
\hat{n}_\mathrm{out}^{(1)}-\hat{n}_\mathrm{out}^{(2)}=r\sqrt{2}\hat{x}(\varphi),
\label{Distinction}
\end{equation}
where
\begin{equation}
\hat{x}(\varphi)=\frac{1}{\sqrt{2}}\left(\hat{b}_{\mathrm{out};
0}^{\dag}e^{i\varphi}+\hat{b}_{\mathrm{out};
0}e^{-i\varphi}\right)
\end{equation}
is the quadrature operator of the CAOM. This means that the AOM does
not affect in cascaded homodyning.

Let $p(x;\alpha, \eta)$ be the phase-averedged quadrature
distribution measured with the shifted amplitude $\alpha$ and the
efficiency $\eta$ given by Eqs.~(\ref{Amplitude}) and
(\ref{EtaOverall}), respectively. Utilizing the results of
Refs.~\cite{Kis, Richter}, one can write the $s$-parametrized
phase-spase distribution of the intracavity mode as
\begin{equation}
P_\mathrm{cav}\!\left(\alpha;s\right)=
\int_{-\infty}^{+\infty}\D x\, S\left(x; s,\eta\right)p(x;\alpha, \eta),
\label{ReconstructionCascaded}
\end{equation}
where the sampling function $S\left(x; s,\eta\right)$ has the form
\begin{equation}
S\left(x; s,\eta\right)=\frac{\eta}{\pi\left[\eta(1-s)-1\right]}
\,f_{00}\!
\left(\frac{x}{\sqrt{\eta(1-s)-1}}\right),
\end{equation}
with $f_{00}(x)$ being expressed in terms of the Dawson integral
$F(x)=e^{-x^2}\int_{0}^{x}\D t\, e^{t^2}$ as
\begin{equation}
f_{00}(x)=2-4xF(x).
\end{equation}
One can use Eq.~(\ref{ReconstructionCascaded}) for the
reconstruction of the phase-space distribution of the intracavity
mode. It is worth noting that the reliability of this method is
completely the same as in the case of a free signal field considered
in Ref.~\cite{Kis}, since the influence of the AOM is eliminated by
the technique itself.

\section{Summary and conclusions}
\label{sec6}

For high-Q cavities, unwanted losses such as the losses due to
scattering and/or absorption may be of the same order of magnitude
as the wanted losses due to the fractional transparency of the
coupling mirrors.  When such cavities are used for the generation
and transfer of nonclassical light, it is of great importance to
carefully consider the noise effects caused by all the unwanted
dissipative channels.

In the present paper we have derived a rather simple and intuitive
extension of the standard quantum noise theory in order to include
in the theory unwanted losses in a consistent way.  For this
purpose, we have modeled the cavity losses by additional beam
splitters that are placed in the input and output channels of the
radiation.  We have analyzed the requirements and constraints for a
complete description of the unwanted losses.  Most importantly, such
a model must ensure that the fundamental commutation rules remain
valid, which allows one to study the possibilities of a complete
parametrization of a cavity with unwanted noise.

To illustrate the relevance of a correct and complete
parameterization, we have also considered an example of a degenerate
model.  It shows that, even though unwanted dissipative channels are
included in the model, the situation may resemble that of a cavity
without unwanted noise. For such cavities information about the
relative phase between intracavity and input modes does not exist in
the outgoing field. In fact, combining the intracavity mode and the
input mode in the nonmonochromatic output mode becomes possible due
to the presence of proper losses inside the cavity and the coupling
mirror.

This mode matching effect can be used, for example, for homodyne
measurements of the intracavity mode. Due to the presence of the
additional output mode satisfactory results for large amplitudes can
be obtained only with a large number of sample events. However, by
using cascaded homodyning the influence of the additional output
mode play no role anymore. In this case the applicability of the
method is much better. The proposed scheme has two advantages
compared with standard homodyning of the extracted field. First, the
phase-space distribution is reconstructed in a point, which is
specified by the value of the local-oscillator amplitude. This
implies that one avoids a complicated numerical integration of
experimental data. Second, one may directly use the properties of
the cavity for combining the signal field with the local oscillator,
avoiding additional mode matching by a beam-splitter and the related
losses. Due to the small efficiency of the quantum-state extraction
from a high-$Q$ cavity, these properties can be useful for
increasing the overall efficiency of the scheme and, consequently,
for obtaining more detailed information about the quantum state of
the intracavity mode.

\begin{acknowledgements}
  This work was supported by Deutsche Forschungsgemeinschaft. A.A.S.  and W.V.
  gratefully acknowledge support by the Deutscher Akademischer
  Austauschdienst. A.A.S. also thanks the President of Ukraine for a research
  stipend.
\end{acknowledgements}

\appendix
\section{Symmetrical and asymmetrical beam splitters} \label{app1}

Let us briefly explain some features of the input-output relation
for the two types of beam-splitters which appear in the replacement
schemes: symmetrical and asymmetrical beam splitters. A symmetrical
beam splitter is a four-port device that is described by the SU(2)
group. The corresponding input-output relations can be written as
\begin{align}
&\hat{a}_\mathrm{out}^{(k)}=\T^{(k)}\hat{a}_\mathrm{in}^{(k)}+\R^{(k)}
\hat{b}_\mathrm{in}^{(k)},\\
&\hat{b}_\mathrm{out}^{(k)}=-\R^{(k)\ast}\hat{a}_\mathrm{in}^{(k)}+
\T^{(k)\ast}\hat{b}_\mathrm{in}^{(k)}.
\end{align}
Inverting these equations, we arrive at
\begin{align}
&\hat{a}_\mathrm{in}^{(k)}=\T^{(k)\ast}\hat{a}_\mathrm{out}^{(k)}-
\R^{(k)}\hat{b}_\mathrm{out}^{(k)},\\
&\hat{b}_\mathrm{in}^{(k)}=\R^{(k)\ast}\hat{a}_\mathrm{out}^{(k)}+
\T^{(k)}\hat{b}_\mathrm{out}^{(k)}.
\end{align}
Here, the index $k$ refers to the respective beam splitter. For
example, for the beam splitter $\mathrm{BS}_1$ in the replacement
scheme in Fig.~\ref{fig5}, we have $\hat{a}_\mathrm{in}^{(1)}$ $\!=$
$\!\hat{g}_\mathrm{in}$, $\hat{b}_\mathrm{in}^{(1)}$ $\!=$
$\!\hat{c}_\mathrm{in}^{(1)}$, $\hat{a}_\mathrm{out}^{(1)}$ $\!=$
$\!\hat{d}_\mathrm{in}$, and $\hat{b}_\mathrm{out}^{(1)}$ $\!=$
$\!\hat{c}_\mathrm{out}^{(1)}$. The transmission and reflection
coefficients $\T^{(k)}$ and $\R^{(k)}$, respectively, which satisfy
the condition
\begin{equation}
\bigl|\T^{(k)}\bigr|^2+\bigl|\R^{(k)}\bigr|^2=1
\label{appeq0}
\end{equation}
can be parametrized by three real numbers $\theta^{(k)}$,
$\mu^{(k)}$, and $\nu^{(k)}$ in the form
\begin{align}
&\T^{(k)}=\cos{\theta^{(k)}}e^{i\mu^{(k)}},\label{appeq1}\\
&\R^{(k)}=\sin{\theta^{(k)}}e^{i\nu^{(k)}}.\label{appeq2}
\end{align}
It is clear that the determinant of the transform matrix is equal to
$1$ in this case---the case of a symmetrical beam splitter.

In the case of an asymmetrical beam splitter, the determinant is an
arbitrary phase multiplier. In fact, this means that this multiplier
should be included in the input-output relations, which then read
\begin{align}
\label{A8}
&\hat{a}_\mathrm{out}^{(k)}=e^{i\varphi^{(k)}}\T^{(k)}\hat{a}_\mathrm{in}^{(k)}
+e^{i\varphi^{(k)}}\R^{(k)}\hat{b}_\mathrm{in}^{(k)},\\\label{A9}
&\hat{b}_\mathrm{out}^{(k)}=-\R^{(k)\ast}\hat{a}_\mathrm{in}^{(k)}+
\T^{(k)\ast}\hat{b}_\mathrm{in}^{(k)}.
\end{align}
This is also a unitary transformation,
however a U(2)-group transformation.
Inverting Eqs.~(\ref{A8}) and (\ref{A9}) yields
\begin{align}
&\hat{a}_\mathrm{in}^{(k)}=e^{-i\varphi^{(k)}}\T^{(k)\ast}
\hat{a}_\mathrm{out}^{(k)}-\R^{(k)}\hat{b}_\mathrm{out}^{(k)},\\
&\hat{b}_\mathrm{in}^{(k)}=e^{-i\varphi^{(k)}}\R^{(k)\ast}
\hat{a}_\mathrm{out}^{(k)}+\T^{(k)}\hat{b}_\mathrm{out}^{(k)}.
\end{align}
The quantities $\T^{(k)}$ and $\R^{(k)}$ again satisfy the condition
(\ref{appeq0}) and can be parametrized according to
Eqs.~(\ref{appeq1}) and (\ref{appeq2}). Clearly the transformation
matrix depends on the additional parameter $\varphi^{(k)}$ and
hence, the resulting number of independent parameters, describing an
asymmetrical beam-splitter is equal to $4$.

\section{Quantum Langevin equation and input-output relation} \label{app2}

Let us start with the derivation of the quantum Langevin equation
(\ref{rs25}). Utilizing the input-output relation (\ref{rs4}) as
well as the input-output relations for each beam-splitter in
Fig.~\ref{fig5} (see Appendix \ref{app1}), we have to first express
the operator $\hat{d}_\mathrm{in}(t)$ in terms of the operators
$\hat{b}_\mathrm{in}(t)$, $\hat{c}^{(1)}_\mathrm{in}(t)$,
$\hat{c}^{(2)}_\mathrm{in}(t)$, and $\hat{a}_\mathrm{cav}(t)$. For
this purpose we apply the input-output relation for the first beam
splitter
\begin{equation}
\hat{d}_\mathrm{in}(t)=\T^{(1)}\hat{g}_\mathrm{in}(t)+
\R^{(1)}\hat{c}^{(1)}_\mathrm{in}(t),\label{rs5}
\end{equation}
and then we find an appropriate expression for the operator
$\hat{g}_\mathrm{in}(t)$. Further, we apply the input-output
relations for the beam splitters in Fig.~\ref{fig5} starting from
the third one and moving clockwise in the loop. The formal sequence
of operations looks as follows.
\begin{enumerate}
\item Substitute $\hat{g}_\mathrm{out}(t)$ from the
  input-output relation
\begin{equation}
\hat{g}_\mathrm{out}(t)=\T^{(2)}\hat{d}_\mathrm{out}(t)+
\R^{(2)}\hat{c}^{(2)}_\mathrm{in}(t)\label{rs14}
\end{equation}
for the second beam splitter into the input-output relation
\begin{equation}
\hat{g}_\mathrm{in}(t)=-\R^{(3)\ast}\hat{g}_\mathrm{out}(t)+
\T^{(3)\ast}\hat{b}_\mathrm{in}(t) \label{rs21}
\end{equation}
for the third beam splitter.
\item Substitute in the resulting equation $\hat{d}_\mathrm{out}(t)$ from
the input-output
relations for the cavity, Eq.~(\ref{rs4}).
\item Substitute in the resulting equation $\hat{d}_\mathrm{in}(t)$ from
Eq.~(\ref{rs5}).
This leads to
\begin{align}
\label{rs23}
&\hat{g}_\mathrm{in}(t)
=-\R^{(3)\ast}\T^{(2)}\sqrt{\gamma}\,
\hat{a}_\mathrm{cav}(t)+
\T^{(3)\ast}\hat{b}_\mathrm{in}(t)
\nonumber\\&\hspace{8ex}
+ \T^{(2)}\R^{(1)}\R^{(3)\ast}\hat{c}_\mathrm{in}^{(1)}(t)-
\R^{(2)}\R^{(3)\ast}\hat{c}_\mathrm{in}^{(2)}(t)
\nonumber\\&\hspace{8ex}
+\R^{(3)\ast}\T^{(1)}\T^{(2)}\hat{g}_\mathrm{in}(t)
,
\end{align}
which
contains
the operator $\hat{g}_\mathrm{in}(t)$ in
both sides of the equation.
\item Resolve Eq.~(\ref{rs23})
to find $\hat{g}_\mathrm{in}(t)$,
\end{enumerate}
\begin{widetext}
\begin{align}
\label{rs24}
\hat{g}_\mathrm{in}(t)=
&-\sqrt{\gamma}\,\frac{\R^{(3)\ast}\T^{(2)}}
{1-\R^{(3)\ast}\T^{(1)}\T^{(2)}}\,\hat{a}_\mathrm{cav}(t)
+\frac{\T^{(3)\ast}}
{1-\R^{(3)\ast}\T^{(1)}\T^{(2)}}\,\hat{b}_\mathrm{in}
(t)\nonumber\\[1ex]
&+\frac{\T^{(2)}\R^{(1)}\R^{(3)\ast}}
{1-\R^{(3)\ast}\T^{(1)}\T^{(2)}}\,\hat{c}_\mathrm{in}^{(1)}
(t) -\frac{\R^{(2)}\R^{(3)\ast}}
{1-\R^{(3)\ast}\T^{(1)}\T^{(2)}}\,\hat{c}_\mathrm{in}^{(2)}
(t).
\end{align}
\end{widetext}%

Inserting Eq.~(\ref{rs24}) into Eq.~(\ref{rs5}) and then
substituting the result into Eq.~(\ref{rs3}), we obtain the quantum
Langevin equation (\ref{rs25}). Next, combining Eq.~(\ref{rs24})
with the (inverse) input-output relation
\begin{equation}
\hat{b}_\mathrm{in}(t)
=e^{-i\varphi^{(3)}}\R^{(3)\ast}\hat{b}_\mathrm{out}(t)
+\T^{(3)}\hat{g}_\mathrm{in}(t) \label{rs26}
\end{equation}
for the third (asymmetrical) beam splitter in Fig.~\ref{fig5}, we
arrive at the sought input-output relation (\ref{rs27}).

\end{document}